\documentclass[11pt]{article}
\usepackage{amsmath,amssymb,color,graphics,epsfig,cite}

\usepackage{amsfonts}

\newcommand{\hoch}[1]{$\, ^{#1}$}

\newcommand{\be}{\begin{equation}}
\newcommand{\ee}{\end{equation}}
\newcommand{\bea}{\setlength\arraycolsep{2pt} \begin{eqnarray}}
\newcommand{\eea}{\end{eqnarray}}
\newcommand{\nn}{\nonumber}

\def\ft#1#2{{\textstyle{\frac{\scriptstyle #1}{\scriptstyle #2} } }}
\def\fft#1#2{{\frac{#1}{#2}}}

\def\0{{\sst{(0)}}}
\def\1{{\sst{(1)}}}
\def\2{{\sst{(2)}}}
\def\3{{\sst{(3)}}}
\def\4{{\sst{(4)}}}
\def\5{{\sst{(5)}}}
\def\6{{\sst{(6)}}}
\def\7{{\sst{(7)}}}
\def\8{{\sst{(8)}}}
\def\sst#1{{\scriptscriptstyle #1}}
\def\oneone{\rlap 1\mkern4mu{\rm l}}

\begin{document}

\begin{center}
{\Large {\bf Boosted Rotating Dyonic Strings in Salam-Sezgin Model }}

\vspace{20pt}

{\large Liang Ma\hoch{1}, Yi Pang\hoch{1} and H. L\"u\hoch{1,2}}

\vspace{10pt}

{\it \hoch{1}Center for Joint Quantum Studies and Department of Physics,\\
School of Science, Tianjin University, Tianjin 300350, China }

\bigskip

{\it \hoch{2}Joint School of National University of Singapore and Tianjin University,\\
International Campus of Tianjin University, Binhai New City, Fuzhou 350207, China}

\vspace{40pt}

\underline{ABSTRACT}
\end{center}

We show that the bosonic sector of the $N=(1,0),\, 6D$ Salam-Sezgin gauged  supergravity model possesses a $T$-duality symmetry upon a circle reduction to $D=5$. We then construct a simple magnetic rotating string solution with two equal angular momenta. Applying the $T$-duality transformation to this solution, we obtain the general boosted rotating dyonic black string solutions whose global structures and thermodynamic quantities are also analyzed. Owing to the fact that the solutions are not asymptotically flat, we find that there are two distinct globally-different non-extremal solutions with two different sets of thermal dynamic variables, with both satisfying the thermodynamic first law and the corresponding Small relations. However, their BPS limit becomes the same and we show that it preserves one quarter of supersymmetry by directly solving the corresponding Killing spinor equations.

\vfill{\footnotesize liangma@tju.edu.cn\ \ \ pangyi1@tju.edu.cn\ \ \ mrhonglu@gmail.com
 }

\thispagestyle{empty}
\pagebreak

\tableofcontents
\addtocontents{toc}{\protect\setcounter{tocdepth}{2}}

\newpage

\section{Introduction}

Six-dimensional supergravities admit many different gaugings and diverse vaccuum solutions. For instance, the gauged $N=(2,2)$ supergravity model was achieved in \cite{Bergshoeff:2007ef}. Unlike maximum supergravities in four, five and seven dimensions, the six-dimensional gauged $N=(2,2)$ supergravity model does not admit a maximally supersymmetric anti-de Sitter (AdS) vacuum. On the other hand, the gauged $N=(1,1)$ supergravity constructed by Romans \cite{Romans:1985tw} does possess supersymmetric AdS$_6$ vacuum solutions, thus enjoying interesting applications in holography. The 6D Romans' theory can be obtained from massive type IIA supergravity from consistent Pauli sphere reduction \cite{Cvetic:1999un} and the corresponding AdS$_6$ vacuum can be interpreted as a D4/D8-brane configuration \cite{Brandhuber:1999np}. Gauged $N=(1,1)$ supergravity with  general matter couplings was obtained in \cite{Andrianopoli:2001rs}.

In this paper, we study the simplest 6D gauged supergravity model with (1,0) supersymmetry, namely the Salam-Sezgin model \cite{Salam:1984cj}. From ungauged supergravity point of view, the minimum model contains a tensor and an abelian vector multiplet, in addition to the minimum supergravity multiplet. Extensions of this model by coupling to more matter multiplets in a way that is free of local anomalies were proposed in \cite{Randjbar-Daemi:1985tdc,Avramis:2005hc,
Avramis:2005qt}. (See also \cite{Pang:2020rir,Becker:2023zyb}.) One intriguing feature of the Salam-Sezgin model is that it admits a half-supersymmetric Minkowski$_4\times S^2$ vacuum, where the $S^2$ is supported by the magnetic dipole charge carried by the $U(1)$ vector field, together with the dilaton potential. It was later found that such a vacuum can also emerge in some variant $N=(1,1)$ gauged supergravities \cite{Kerimo:2003am, Kerimo:2004md}. Subsequently, a large class of gauged supergravities with Minkowski$\times$sphere vacua were classified in \cite{Kerimo:2004qx}.

Another intriguing feature of the Salam-Sezgin model is that for a vacuum to preserve supersymmetry, the supertransformation of the gaugino leads to
\be
(e^{\fft14\varphi} F_{\mu\nu}\Gamma^{\mu\nu} - 8{\rm i} g e^{-\fft14 \varphi})\epsilon=0\,,
\ee
which implies that preserving any amount of supersymmetry requires non-trivial $U(1)$ flux $F_{\mu\nu}$.  By contrast, in ungauged theory with $g=0$, it is preferable to set $F_{\mu\nu}=0$ for the construction of BPS solutions.  Indeed, $\ft14$-BPS dyonic string solution in Salam-Sezgin model \cite{Gueven:2003uw} involves a magnetic dipole charge of $F_{\mu\nu}$. Its non-extremal generalization was recently constructed in \cite{Ma:2023tcj} where the string charge lattice was also analyzed.

In this paper, we shall add angular momentum to the dyonic string and study their BPS limit, generalizing the results of \cite{Gueven:2003uw,Ma:2023tcj}. In six dimensions, a string solution has 4-dimensional transverse space so that the rotation group is $SO(4)$ with two independent orthogonal rotations. However, owing to the necessity of involving the magnetic dipole charge, the construction of the rotating solutions becomes more subtle. For the static solutions, the magnetic dipole charge has the effect of squashing $U(1)$ fibre over the $S^2$ base of the 3-sphere in the transverse direction. We therefore consider only ``two equal'' angular momenta $J_a=J_b$ so that the rotation occurs only in the $U(1)$ fibre direction, while the $S^2$ base space is preserved. This greatly simplify the construction, but it can be still rather complicated if the target solution carries both the electric and magnetic string charges associated with the 3-form field strength, magnetic dipole charge of the 2-form field strength as well as both angular and boosted linear momenta.

Our breakthrough comes from the observation that the scalar potential of Salam-Sezgin model takes exact the same form as the conformal anomaly term in non-critical strings \cite{Callan:1985ia}. It has the consequence that the T-duality symmetry at the level of supergravity Lagrangian is still preserved in the Salam-Sezgin model.  We find that the five-dimensional theory from $S^1$ reduction of the Salam-Sezgin model has a nonlinearly realized $SO(2,1)/SO(2)$ coset symmetry, under which the three abelian vector fields, namely the Kaluza-Klein vector, winding vector from $B_{\mu\nu}$ and the vector descending directly from the 6D $U(1)$ gauge field, form a triplet.  We give explicit global symmetry transformations of the $SO(2,1)$ acting on various fields. With this $T$-duality symmetry, we are able to construct the boosted rotating dyonic string solutions. In a typical ungauged supergravity, any Ricci-flat metric is a vacuum solution. We can thus start with the known neutral Myers-Perry rotating black hole metric as a seed solution and perform the appropriate Kaluza-Klein reduction. The global symmetry of the reduced theory can be used to generate more solutions. Upon lifting back to the original dimension, one can obtain charged rotating black holes. However, Ricci-flat metrics are not solutions in the Salam-Sezgin model; therefore, the usual Myers-Perry metric cannot be used as a seed solution. Furthermore, the involvement of the magnetic dipole charges make the construction of a simpler seed solution even more complicated. Nevertheless, we overcome this problem and obtain the seed solution by direct construction. Another consequence of that Ricci-flat metrics are not solutions of the Salam-Sezgin model is that the black holes are not asymptotic to flat Minkowski spacetime. We therefore do not have a fiducial spacetime for fixing the scaling symmetry in the time direction. We find that this can lead to two globally-different solutions with two distinct sets of thermodynamic variables, but both satisfying the first law of thermodynamics and corresponding Small relations.

The paper is organized as follows.  In section 2, we analyse the T-duality symmetry of Salam-Sezgin model reduced on $S^1$. We obtain the symmetry transformation rule and manifestly $SO(2,1)$-invariant form in both Einstein and string frames. In section 3, we construct magnetic rotating string solution as a seed solution and obtain the general boosted rotating dyonic solutions.  We find two globally different non-extremal solutions with different set of thermodynamic rules. However, in section 4, we show that they give the same BPS limits and we obtain the Killing spinors. We conclude our paper in section 5.

\section{T-duality symmetry of Salam-Sezgin model}

The Salam-Sezgin model in six dimensions is the minimum $N=(1,0)$ gauged supergravity. The bosonic sector consists of the metric and matter fields $(B_\2, A_\1, \varphi)$. The Lagrangian is \cite{Salam:1984cj}
\bea
\mathcal{L}_6&=& R-\frac{1}{4}\left(\partial\varphi\right)^2-\frac{1}{12}e^\varphi H_\3^2-\frac{1}{4}e^{\frac{1}{2}\varphi}F_\2^2-8g^2e^{-\frac{1}{2}\varphi},\nn\\
F_{\2}&=&dA_{\1}\,,\qquad H_\3=dB_\2+\frac{1}{2}F_{\2}\wedge A_{\1}\,.\label{Salam-Sezgin-L}
\eea
For convenience we omit the universal factor $\sqrt{-g}$ throughout the paper. Note that the subscript $(n)$ denotes the associated quantity is an $n$-form. This bosonic sector resembles the noncritical bosonic string theory with the conformal anomalous term that admits pseudo-supersymmetry \cite{Lu:2011ku}.

In fact, the bosonic Lagrangian \eqref{Salam-Sezgin-L}, on its own, can also be obtained from the seven-dimensional noncritical string theory via the $S^1$ Kaluza-Klein reduction, after setting the Kaluza-Klein and winding vectors equal to $A_\1$, which allows one to truncate out consistently one combination of the two dilatonic scalars. Since the conformal anomaly in string theory preserves the T-duality, the seven-dimensional noncritical string theory reduced on two torus will have $SO(2,2)$ T-duality global symmetry. Setting one pair of Kaluza-Klein and winding modes equal reduces the global symmetry to the diagonal $SO(2,1)\sim SL(2,\mathbb R)$. Thus the global symmetry of the Salam-Sezgin model reduced on $S^1$ should have $SO(2,1)$ global symmetry. We next derive the full set of symmetry transformation rules that is useful for our construction of the rotating dyonic solution.

\subsection{Kaluza-Klein reduction to $D=5$}

The standard Kaluza-Klein reduction ansatz of the Salam-Sezgin model \eqref{Salam-Sezgin-L} on $S^1$ associated with $x$ coordinate is
\bea
ds_6^2&=&e^{\frac{1}{2\sqrt{3}}\phi}d\tilde{s}_5^2+e^{-\frac{\sqrt{3}}{2}\phi}
(dx+\tilde{\mathcal{A}}_{(1)})^2\,,\cr
B_{\2}&=&\tilde{B}_{\2}'+\tilde{\mathcal{B}}'_{\1}\wedge dx,\quad A_{\1}=\tilde{A}_{\1}'+
\psi dx\,.\label{reduction ansatz}
\eea
For later purpose, it is advantageous to redefine the scalar field
\bea
\varphi-\frac{1}{\sqrt{3}}\phi=\frac{2}{\sqrt{3}}\phi_1\,,\qquad
\frac{1}{\sqrt{3}}\varphi+\phi=\frac{2}{\sqrt{3}}\phi_2\,,
\eea
and the form fields
\bea
\tilde{A}_{\1}'&=&\tilde{A}_{\1}+\psi \tilde{\mathcal{A}}_{\1}\,,\qquad
\tilde{\mathcal{B}}_{\1}'=\tilde{\mathcal{B}}_{\1}+\frac{1}{2}\psi \tilde{A}_{\1}\,,\cr
\tilde{B}_{(2)}'&=&\tilde{B}_{(2)}+\frac{1}{2}\tilde{\mathcal{B}}_{(1)}
\wedge\tilde{\mathcal{A}}_{(1)}+\frac{1}{2}\psi\tilde{A}_{\1}\wedge
\tilde{\mathcal{A}}_{\1}\,.
\eea
The reduced five-dimensional theory in the Einstein frame becomes
\bea
\mathcal{L}_5^\mathrm{E}&=&\tilde{R}_5-\frac{1}{4}\left(\nabla\phi_1\right)^2
-\frac{1}{4}\left(\nabla\phi_2\right)^2-\frac{1}{4}e^{\frac{1}{\sqrt{3}}\phi_1}
\left(
e^{-\phi_2}\tilde{\mathcal{F}}^{0,2}_\2+e^{\phi_2}
\underline{\tilde{\mathcal{H}}}_\2^2+\tilde{F}_\2^2
\right)\cr
&&-\frac{1}{12}e^{\frac{2}{\sqrt{3}}\phi_1}\underline{\tilde{H}}_\3^2
-\frac{1}{2}e^{\phi_2}(\nabla\psi)^2
-8g^2e^{-\frac{1}{\sqrt{3}}\phi_1}.\label{5E Einstein action}
\eea
Here we define some shorthand notations
\bea
\tilde{F}_{\2}&=&\tilde{F}_{\2}^0+\psi\tilde{\mathcal{F}}^0_{\2}\,,\qquad
\underline{\tilde{\mathcal{H}}}_{\2}=\tilde{\mathcal{H}}^0_{\2}
+\psi\tilde{F}_{\2}^0+\frac{1}{2}\psi^2\tilde{\mathcal{F}}^0_{\2}\,,\cr
\underline{\tilde{H}}_{\3}&=&\tilde{H}^0_{\3}-\frac{1}{2}\tilde{\mathcal{H}}^0_{\2}
\wedge\tilde{\mathcal{A}}_{\1}
-\frac{1}{2}\tilde{\mathcal{B}}_{(1)}\wedge\tilde{\mathcal{F}}^0_{\2}+
\frac{1}{2}\tilde{F}_{\2}^0\wedge\tilde{A}_{\1}\,.\label{shorthand notation}
\eea
Quantities with a superscript ``0" denote close field strengths without Kaluza-Klein modifications, {\it i.e.}
\be
\tilde{F}_{\2}^0= d\tilde{A}_{\1}\,,\qquad \tilde{\mathcal{F}}^0_{\2}=d\tilde{\mathcal{A}}_{\1}\,,\qquad \tilde{\mathcal{H}}^0_{\2}=d\tilde{\mathcal{B}}_{\1}\,,\qquad \tilde{H}^0_{\3}=d\tilde{B}_{\2}\,.
\ee

\subsection{Global symmetry}

At the first sight, the two scalars $(\phi_2, \psi)$ form a complex scalar describing the coset of $SL(2,\mathbb R)/SO(2)$ and $(\tilde{\mathcal{F}}_\2^0,\tilde{\mathcal{H}}_\2^0)$ form a doublet under
the $SL(2,\mathbb R)$ global symmetry, but this does not fit the Kaluza-Klein modifications of the field strengths. Instead, we should treat the scalar pair as the coset structure of the isomorphic $SO(2,1)/SO(2)$, with the three vector fields forming a triplet under the $SO(2,1)$. To make this idea concrete, we introduce the 3$\times$3 Cartan generator $H$, and the upper and lower triangular root generators $E_\pm$
\bea
H=
\begin{bmatrix}
  1 & 0 & 0 \\
  0 & 0 & 0 \\
  0 & 0 & -1 \\
\end{bmatrix},\qquad
E_+=\begin{bmatrix}
 0 & 1 & 0  \cr
 0 & 0 & 1  \cr
 0 & 0 & 0
\end{bmatrix},\qquad
E_-=\begin{bmatrix}
 0 & 0 & 0  \cr
 1 & 0 & 0  \cr
 0 & 1 & 0
\end{bmatrix}.\label{HE}
\eea
They satisfy the algebra
\bea
[H,E_\pm]=\pm E_-\,,\qquad [E_+,E_-]=H\,.
\eea
We can now parameterise the coset $\mathcal{V}$
\bea
\mathcal{V}=e^{\frac{1}{2}\phi_2H}e^{\psi E_+}=\begin{bmatrix}
 e^{\frac{1}{2}\phi_2} & \psi e^{\frac{1}{2}\phi_2} & \frac{1}{2}\psi^2e^{\frac{1}{2}\phi_2}  \cr
 0 & 1 & \psi  \cr
 0 & 0 & e^{-\frac{1}{2}\phi_2}
\end{bmatrix},
\eea
and define $\mathcal{M}=\mathcal{V}^T\mathcal{V}$. The kinetic term of $(\phi_2,\psi)$ can be expressed in a standard way by $\mathcal{M}$, namely
\bea
\frac{1}{8}\mathrm{Tr}[(\partial\mathcal{M}^{-1})(\partial\mathcal{M})]
=-\frac{1}{4}\left(\nabla\phi_2\right)^2-\frac{1}{2}e^{\phi_2}(\nabla\psi)^2.
\eea
We define a 1-form vector field triplet and its field strength
\be
\mathbb{B}_\mu=\begin{bmatrix}
 \tilde{\mathcal{B}}_\mu &   \\
 \tilde{A}_\mu &   \\
 \tilde{\mathcal{A}}_\mu &
\end{bmatrix}\,,\qquad
\mathbb{H}_{\mu\nu}=\begin{bmatrix}
 \tilde{\mathcal{H}}_{\mu\nu}^0 &   \\
 \tilde{F}_{\mu\nu}^0 &   \\
 \tilde{\mathcal{F}}_{\mu\nu}^0 &
\end{bmatrix}.
\ee
The corresponding kinetic terms from an invariant bilinear construction are
\be
\mathbb{H}_{\mu\nu}^T\mathcal{M}\mathbb{H}^{\mu\nu}
=e^{-\phi_2}\tilde{\mathcal{F}}_\2^{0,2}+e^{\phi_2}
\underline{\tilde{\mathcal{H}}}^2_\2+\tilde{F}^2_\2.
\ee
The 2-form potential and its 3-form field strength are singlets under the $SO(2,1)$ global symmetry. In order to see that its Kaluza-Klein modification is indeed a singlet, we define a matrix $\eta=\eta^{-1}$
\be
\eta=\begin{bmatrix}
 0 & 0 & -1  \cr
 0 & 1 & 0  \cr
 -1 & 0 & 0
\end{bmatrix},\qquad \mathcal{V}^T\eta\mathcal{V}=\eta\,.
\ee
We can now express the three-form field strength manifestly as a singlet:
\bea
\underline{\tilde{H}}_{\3}&=&d\tilde{B}_{(2)}+\frac{1}{2}\mathbb{H}_{(2)}^T
\wedge\left(\eta\mathbb{B}_{(1)}\right).
\eea
With these, we can write the five-dimensional reduced theory \eqref{5E Einstein action} in a manifestly invariant form, {\it i.e.},
\bea
\mathcal{L}_5^\mathrm{E}&=&\tilde{R}_5-\frac{1}{4}\left(\nabla\phi_1\right)^2
+\frac{1}{8}\mathrm{Tr}[(\partial\mathcal{M}^{-1})(\partial\mathcal{M})]\cr
&&
-\frac{1}{4}e^{\frac{1}{\sqrt{3}}\phi_1}\mathbb{H}_{\mu\nu}^T\mathcal{M}
\mathbb{H}^{\mu\nu}
-\frac{1}{12}e^{\frac{2}{\sqrt{3}}\phi_1}\underline{\tilde{H}}_\3^2
-8g^2e^{-\frac{1}{\sqrt{3}}\phi_1},\cr
\underline{\tilde{H}}_{\3}&=&d\tilde{B}_{(2)}
+\frac{1}{2}\mathbb{H}_{(2)}^T\wedge\left(\eta\mathbb{B}_{(1)}\right).
\label{5E Einstein action rewrite}
\eea
It can be easily seen that the theory is invariant under the general $SO(2,1)$ global transformation $S$
\be
\mathcal{M}\rightarrow\mathcal{M}'=S^T\mathcal{M}S\,,\qquad \mathbb{B}_{(1)} \rightarrow \mathbb{B}_{(1)}'=S^{-1}\mathbb{B}_{(1)}\,,\qquad
S^T\eta S=\eta\,.
\ee
We can parameterize $S$ by
\be
S=S_1S_2S_3\,,\qquad S_1=e^{t_1 H}\,,\qquad S_2 = e^{t_2 E_+}\,,\qquad
S_3=e^{t_3 E_-}\,.
\ee
Specifically, we have
In finite version, the transformation matrix $S$ is given by
\bea
S_1=
\begin{bmatrix}
  e^{t_1} & 0 & 0 \\
  0 & 1 & 0 \\
  0 & 0 & e^{-t_1} \\
\end{bmatrix},\quad
S_2=\begin{bmatrix}
 1 & t_2 & \frac{t_2}{2}  \cr
 0 & 1 & t_2  \cr
 0 & 0 & 1
\end{bmatrix},\quad
S_3=\begin{bmatrix}
 1 & 0 & 0  \cr
 t_3 & 1 & 0  \cr
 \frac{t_3^2}{2} & t_3 & 1
\end{bmatrix}.
\eea
We can now give the explicit transformation rules of scalar fields $\{\phi,\psi\}$ and form fields $\{\tilde{\mathcal{B}}_\mu,\tilde{A}_\mu, \tilde{\mathcal{A}}_\mu\}$ from
\bea
\mathcal{M}'(\phi',\psi')=S^T\mathcal{M}(\phi,\psi)S,\quad
\mathbb{B}_{(1)}'(\tilde{\mathcal{B}}_\mu',
\tilde{A}_\mu', \tilde{\mathcal{A}}_\mu')=S^{-1}\mathbb{B}_{(1)}(\tilde{\mathcal{B}}_\mu,
\tilde{A}_\mu, \tilde{\mathcal{A}}_\mu).
\eea
They are
\begin{itemize}
\item $S_1$: \bea
\phi&\rightarrow&\phi'=\phi+2t_1,\quad \psi\rightarrow\psi'=e^{-t_1}\psi\,,\cr
\tilde{\mathcal{B}}_\mu&\rightarrow&
\tilde{\mathcal{B}}_\mu'=e^{-t_1}\tilde{\mathcal{B}}_\mu\,,\qquad
\tilde{A}_\mu\rightarrow\tilde{A}_\mu'=\tilde{A}_\mu\,,\qquad
\tilde{\mathcal{A}}_\mu\rightarrow\tilde{\mathcal{A}}_\mu'=e^{t_1}
\tilde{\mathcal{A}}_\mu\,.
\eea
\item $S_2$: \bea
\phi&\rightarrow&\phi'=\phi,\quad \psi\rightarrow\psi'=\psi+t_2,\cr
\tilde{\mathcal{B}}_\mu&\rightarrow&
\tilde{\mathcal{B}}_\mu'=\tilde{\mathcal{B}}_\mu-t_2\tilde{A}_\mu+\frac{t_2^2}{2}
\tilde{\mathcal{A}}_\mu\,,\quad
\tilde{A}_\mu\rightarrow\tilde{A}_\mu'=\tilde{A}_\mu-t_2\tilde{\mathcal{A}}_\mu\,,
\quad
\tilde{\mathcal{A}}_\mu\rightarrow\tilde{\mathcal{A}}_\mu'=\tilde{\mathcal{A}}_\mu\,.
\eea
\item $S_3$: \bea
\phi&\rightarrow&\phi'=-\phi+2 \log \left(\frac{t_3^2}{2}+e^{\phi } \left(\frac{t_3 \psi }{2}+1\right)^2\right),\quad \psi\rightarrow\psi'=\frac{2 t_3 \left(\psi ^2 e^{\phi }+2\right)+4 \psi  e^{\phi }}{e^{\phi } \left(t_3 \psi +2\right){}^2+2 t_3^2},\cr
\tilde{\mathcal{B}}_\mu&\rightarrow&
\tilde{\mathcal{B}}_\mu'=\tilde{\mathcal{B}}_\mu\,,\quad
\tilde{A}_\mu\rightarrow\tilde{A}_\mu'=\tilde{A}_\mu-t_3\tilde{\mathcal{B}}_\mu\,,
\quad
\tilde{\mathcal{A}}_\mu\rightarrow\tilde{\mathcal{A}}_\mu'
=\tilde{\mathcal{A}}_\mu-t_3\tilde{A}_\mu
+\frac{t_3^2}{2}\tilde{\mathcal{B}}_\mu\,.
\eea
\end{itemize}
Note that the dilaton $\phi_1$ generates a constant shift $\mathbb R$ symmetry, so that the total global symmetry is $SO(2,1)\times \mathbb R\sim GL(2,\mathbb R)$. Note that if we set the Kaluza-Klein and winding vectors equal, we can consistently truncate out the $SO(2,1)/SO(2)$ scalar coset. This is analogous the $D=7$ to $D=6$ reduction, where a vector multiplet can be consistently truncated out, commented above section 2.1.

\subsection{String frame}

It is also instructive to discuss the global symmetry in the string frame. Under the conformal transformation
\be
g_{\mu\nu}^\mathrm{E}=e^{\frac{1}{2}\varphi}g_{\mu\nu}^\mathrm{S}\,,
\ee
the Salam-Sezgin model \eqref{Salam-Sezgin-L} becomes
\bea
\mathcal{L}_6^\mathrm{S}&=&e^\varphi\left(R_6\, {\star_6 \oneone}-\frac{1}{2}{\star_6H_{\3}}\wedge H_{\3}-\frac{1}{2}{\star_6F_{\2}}\wedge F_{\2}+\star_6d\varphi\wedge d\varphi-8g^2{\star_6 \oneone}\right).\label{d6string}
\eea
We can now see clearly that the $g^2$ indeed appears as if it is the conformal anomaly term in a noncritical string. We consider the circle reduction
\bea
ds_6^2&=&d\tilde{s}_5^2+e^{-\phi}(dx+\tilde{\mathcal{A}}_{\1})^2,\cr
B_{\2}&=&\tilde{B}_{\2}'+\tilde{\mathcal{B}}_{\1}'\wedge dx\,,\qquad A_{\1}=\tilde{A}_{\1}'+\psi dx\,,\label{d6stringred}
\eea
and set $\varphi=\frac{1}{2}\phi-2\Phi$. We obtain the five-dimensional theory in the string frame, namely
\bea
\mathcal{L}_5^\mathrm{S}&=&e^{-2\Phi}\left[
\tilde{R}_5+4(\nabla\Phi)^2+\frac{1}{8}\mathrm{Tr}
[(\partial\mathcal{M}^{-1})(\partial\mathcal{M})]
-\frac{1}{12}\underline{\tilde{H}}_\3^2-\frac{1}{4}\mathbb{H}_{\mu\nu}^T\mathcal{M}
\mathbb{H}^{\mu\nu}-8g^2\right],\cr
\underline{\tilde{H}}_{\3}&=&d\tilde{B}_{(2)}+\frac{1}{2}\mathbb{H}_{(2)}^T\wedge
\left(\eta\mathbb{B}_{(1)}\right).\label{5S2}
\eea
It is manifestly invariant under $SO(2,1)$. Note that we obtained the above equations by appropriate conformal transformation to convert the calculations in section 2.2 in the Einstein frame to the string frame. We also verified that \eqref{5S2} can indeed be obtained directly from \eqref{d6string} on the reduction ansatz \eqref{d6stringred}. Since many black hole thermodynamic expressions are developed in the Einstein frame, we find it is advantageous to construct and study the rotating black holes in the Einstein frame.

\section{Rotating dyonic string solution}
\label{sec:sol1}

A direct construction of rotating solutions can be a formidable task in the Salam-Sezgin model, when all the fields will be necessarily turned on. In supergravities, one typically adopts the solution generating technique that utilizes the global symmetry in the Kaluza-Klein reduced theory.  Notable examples of such construction include the Sen \cite{Sen:1992ua} and Cveti\v c-Youm solutions \cite{Cvetic:1996xz}. Such global symmetries are typically broken in gauged supergravities, making the construction much more subtle, e.g.~\cite{Chong:2005hr,Wu:2011gq}

Fortunately, as we have seen in the previous section, the T-duality survives in Salam-Sezgin model despite of the gauging. However, there is still an extra subtlety arising from the gauging. For the ungauged theory, we can start with a neutral rotating black string, which is a direct product a line and five-dimensional Myers-Perry black hole, as the seed solution and generate the charged ones by the solution-generating technique. In the gauged theory, there is no Ricci-flat vacuum. The only known solutions in literature are the static dyonic strings and their BPS limit \cite{Gueven:2003uw,Ma:2023tcj}. Furthermore, the ``harmonic function'' associated with the magnetic string charge takes the form $H_P\sim P/\rho^2$. The BPS constraint requires that magnetic dipole charge $k$ and the $P$ are constrained by some algebraic relation. This implies that the minimum solution necessarily contains both $(P,k)$ parameters. We therefore construct first the seed solution of rotating strings carrying both $(P,k)$ parameters.

\subsection{Magnetic seed solution}

The $\ft14$-BPS static dyonic string with magnetic dipole charge was constructed in \cite{Gueven:2003uw}. Its non-extremal generalization was obtained in \cite{Ma:2023tcj}. Turning off the electric charge, the two form fields are given by
\be
A_{\1}=-k\cos\theta d\varphi\,,\qquad B_{\2}=-\frac{\mu s_1c_1}{2}(1+\cos\theta) d\varphi\wedge d\psi\,.
\ee
Here we parameterize the magnetic string charge $P$ as $P=\fft12 s_1 c_1$. In this paper, we denote $(s_i, c_i)$ as
\be
s_i = \sinh \delta_i\,,\qquad c_i=\cosh\delta_i\,.
\ee
The metric takes the form
\be
ds^2_{6,\mathrm{E}}=\frac{1}{\sqrt{H_1}}\left(-h_1dt^2+dx^2\right)+\sqrt{H_1}\Big(
\xi_2\frac{d\rho^2}{\Delta_\rho}+\frac{1}{4}\rho^2\left(\sigma_3^2+\xi_1d\Omega_2^2\right)
\Big),
\ee
with $\sigma_3=d\psi+\cos\theta d\varphi$ and  $d\Omega_2^2=d\theta^2 + \sin^2 \theta d\varphi^2$ \cite{Gueven:2003uw,Ma:2023tcj}.  Note that the level surfaces in the four-dimensional transverse is not the round 3-sphere, but the squashed one, described as a squashed $U(1)$ bundle over $S^2$. The rotation thus occurs in the fibre direction. We find that the full magnetic rotating string is
\bea
ds^2_{6,\mathrm{E}}&=&\frac{1}{\sqrt{H_1}}\left(-h_1dt^2+dx^2-h_2dt\sigma_3\right)
+\sqrt{H_1}\Big(\xi_2\frac{d\rho^2}{\Delta_\rho}+\frac{1}{4}\rho^2
\left(\sigma_3^2+\xi_1d\Omega_2^2\right)
\Big),\label{seed1}\\
A_{\1}&=&-k\cos\theta d\varphi,\quad B_{\2}=-\frac{\mu s_1c_1}{2}(1+\cos\theta) d\varphi\wedge d\psi-\frac{\mu a \sqrt{s_1c_1}}{\rho^2}dt\wedge\sigma_3\,,\quad
e^{-\varphi}=H_1\,,\nn
\eea
where the functions $(h_1, h_2, \Delta_\rho, H_1)$ and the two constants $(\xi_1,\xi_2)$ are
\bea
h_1&=&1-\frac{2\mu}{\rho^2},\quad h_2=\frac{2\mu a\sqrt{s_1c_1}}{\rho^2},\quad
\Delta_\rho=1-\frac{2\mu}{\rho^2}+\frac{2\mu a^2}{\rho^4},\quad H_1=\frac{2\mu s_1c_1}{\rho^2}\,,\cr
\xi_1&=&\frac{k^2+\mu s_1c_1}{\mu s_1c_1}\,,\qquad
\xi_2=\left(\xi_1^{-1}-4\mu s_1c_1g^2\right)^{-1}\,.\label{xi1xi2}
\eea

\subsection{Generating the electric string charges}

In the effective theory of strings, there is a usual process of generating the electric string charge. One can perform a Lorenz boost along the string direction and then perform Kaluza-Klein circle reduction and obtain the electrically-charged black hole in one lower dimensions. One can then use the T-duality to map the Kaluza-Klein charge to winding charge and lift the solution black to the original dimensions. We can follow the same procedure on the magnetic seed solution \eqref{seed1}, since we have established that the T-duality is preserved in five dimensions even for the gauged theory.

We start by performing a Lorentz boost
\bea
t\rightarrow c_2t+s_2x\,,\qquad x\rightarrow c_2x+s_2t\,.
\eea
The solution \eqref{seed1} can be expressed in terms of the form that is ready for Kaluza-Klein reduction
\bea
ds^2_{6,\mathrm{E}}&=&\frac{1}{\sqrt{H_1}}\left[-(h_1c_2^2-s_2^2)dt^2-h_2c_2dt\sigma_3
-\frac{s_2^2}{c_2^2-h_1s_2^2}\left(c_2(1-h_1)dt-h_2\frac{\sigma_3}{2}\right)^2\right]\cr
&&+\frac{c_2^2-h_1s_2^2}{\sqrt{H_1}}\left[
dx+\frac{s_2}{c_2^2-h_1s_2^2}\left(c_2(1-h_1)dt-h_2\frac{\sigma_3}{2}\right)
\right]^2\cr
&&+\sqrt{H_1}\Big(
\xi_2\frac{d\rho^2}{\Delta_\rho}+\frac{1}{4}\rho^2\left(\sigma_3^2+
\xi_1d\Omega_2^2\right)
\Big),\cr
B_{\2}&=&-\frac{\mu s_1c_1}{2}(1+\cos\theta) d\varphi\wedge d\psi-\frac{\mu a \sqrt{s_1c_1}c_2}{\rho^2}dt\wedge\sigma_3-\frac{\mu a\sqrt{s_1c_1}s_2}{\rho^2}dx\wedge\sigma_3\,,\cr
A_{\1}&=&-k\cos\theta d\varphi\,,\qquad e^{-\varphi}=H_1\,.\label{boost magnetic}
\eea
The reduced five-dimensional solution is therefore given by
\bea
\tilde{\mathcal{A}}_{(1)}&=&\frac{s_2}{c_2^2-h_1s_2^2}
\left(c_2(1-h_1)dt-h_2\frac{\sigma_3}{2}\right)\,,\qquad e^{-\frac{\sqrt{3}}{2}\phi}=\frac{c_2^2-h_1s_2^2}{\sqrt{H_1}}\,,\cr
d\tilde{s}_5^2&=&\left(\frac{c_2^2-h_1s_2^2}{\sqrt{H_1}}\right)^{\frac{1}{3}}
\Bigg\{\frac{1}{\sqrt{H_1}}\Big[-(h_1c_2^2-s_2^2)dt^2-h_2c_2dt\sigma_3\cr
&&
-\frac{s_2^2}{c_2^2-h_1s_2^2}\left(c_2(1-h_1)dt-h_2\frac{\sigma_3}{2}\right)^2\Big]
+\sqrt{H_1}\Big(
\xi_2\frac{d\rho^2}{\Delta_\rho}+\frac{1}{4}\rho^2
\left(\sigma_3^2+\xi_1d\Omega_2^2\right)\Big)\Bigg\}\,,\cr
\tilde{\mathcal{B}}_{\1} &=& \frac{\mu a \sqrt{s_1c_1}s_2}{\rho^2}\sigma_3\,,\qquad
\tilde{B}_{\2}=-\frac{\mu s_1c_1}{2}(1+\cos\theta) d\varphi\wedge d\psi-\frac{\mu a \sqrt{s_1c_1}c_2}{\rho^2}dt\wedge\sigma_3\,,\cr
\tilde{A}_{\1}&=&-k\cos\theta d\varphi,\qquad \psi=0,\qquad e^{-\varphi}=H_1\,.\label{boost magnetic reduction}
\eea
The scalar $\phi_{1,2}$ combinations are
\bea
\phi_1=\frac{1}{\sqrt{3}}\log{\frac{c_2^2-h_1s_2^2}{H_1^2}},\qquad
\phi_2=-\log{(c_2^2-h_1s_2^2)}\,.
\eea
It is clear that for the reduced solution, we have
\be
\tilde{F}^0_{\2}\wedge\tilde{A}_{\1}=0\,,\quad \psi=0\,.
\ee
Consequently, the field strengths defined in \eqref{shorthand notation} are
\bea
\underline{\tilde{H}}_{(3)}&\sim&d\tilde{B}_{\2}-d\tilde{\mathcal{B}}_{\1}
\wedge\tilde{\mathcal{A}}_{\1}\,,\qquad
\underline{\tilde{\mathcal{H}}}_{\2}\sim d\tilde{\mathcal{B}}_{\1}\,,\qquad
\tilde{\mathcal{F}}_{\2}^0\sim d\tilde{\mathcal{A}}_{\1}\,.
\eea
For this reduced set of fields, the five-dimensional theory \eqref{5E Einstein action} is invariant under the transformation rule
\bea
\tilde{\mathcal{A}}_{\1}&\rightarrow&\tilde{\mathcal{A}}_{\1}'=
\tilde{\mathcal{B}}_{\1}\,,\qquad
\tilde{\mathcal{B}}_{\1}\rightarrow\tilde{\mathcal{B}}_{\1}'=
\tilde{\mathcal{A}}_{\1}\,,\cr
\tilde{B}_{\2}&\rightarrow&\tilde{B}_{\2}'=\tilde{B}_{\2}+
\tilde{\mathcal{A}}_{\1}\wedge\tilde{\mathcal{B}}_{\1}\,,\qquad
\phi_2\rightarrow \phi_2'=-\phi_2\,.\label{duality}
\eea
Under this transformation rule, we obtain a new solution in five dimensions where the electric charge is carried by the winding vector $\tilde {\cal B}_1'$ instead. We then lift the solution back to $D=6$, and we obtain the dyonic string solution
\bea
ds_6'^2 &=&\sqrt{c_2^2-h_1s_2^2}\Bigg\{\frac{1}{\sqrt{H_1}}
\Big[-(h_1c_2^2-s_2^2)dt^2-h_2c_2dt\sigma_3\cr
&&
-\frac{s_2^2}{c_2^2-h_1s_2^2}\left(c_2(1-h_1)dt-h_2\frac{\sigma_3}{2}\right)^2\Big]
+\sqrt{H_1}\Big(
\xi_2\frac{d\rho^2}{\Delta_\rho}+\frac{1}{4}\rho^2
\left(\sigma_3^2+\xi_1d\Omega_2^2\right)
\Big)\Bigg\}\cr
&&+\frac{1}{\sqrt{H_1}\sqrt{c_2^2-h_1s_2^2}}\left(dx+\frac{\mu a\sqrt{s_1c_1}s_2}{r^2}\sigma_3\right)^2,\cr
B_{\2}' &=&-\frac{\mu s_1c_1}{2}(1+\cos\theta) d\varphi\wedge d\psi-\frac{\mu a \sqrt{s_1c_1}c_2}{\rho^2(c_2^2-h_1s_2^2)}dt\wedge\sigma_3\cr
&&+\frac{s_2}{c_2^2-h_1s_2^2}\left(c_2(1-h_1)dt-h_2\frac{\sigma_3}{2}\right)\wedge dx,\cr
A_{\1}'&=&-k\cos\theta d\varphi,\qquad \varphi'=\log{\frac{c_2^2-h_1s_2^2}{H_1}}\,.\label{dyonic 6D}
\eea
We can now remove the prime in \eqref{dyonic 6D} and perform a further boost
\be
t\rightarrow c_3t+s_3x\,,\qquad x\rightarrow c_3x+s_3t\,.
\ee
We finally arrive at the rotating and boosted dyonic string in Salam-Sezgin model:
\bea
ds_6^2&=&\frac{1}{\sqrt{H_1(c_2^2-h_1 s_2^2)}}\Bigg[
\left(c_2^2-h_1 s_2^2\right) \left[-c_2 c_3  h_2 dt\sigma _3+c_3^2 \left(s_2^2-c_2^2 h_1\right)dt^2\right]\cr
&&\!\!\!\!\! -s_2^2\left(c_2 c_3  \left(1-h_1\right)dt-\frac{h_2 }{2}\sigma _3\right)^2
+\left(s_3dt+\frac{ \mu a \sqrt{s_1c_1} s_2 }{\rho^2}\sigma _3\right)^2-\frac{K_{\1}^2}{c_3^2-h_1 s_3^2}
\Bigg]\cr
&&\!\!\!\!\!
+\sqrt{H_1(c_2^2-h_1 s_2^2)}\left(
\xi_2\frac{d\rho^2}{\Delta_\rho}+\frac{1}{4}\rho^2
\left(\sigma_3^2+\xi_1d\Omega_2^2\right)
\right)
+\frac{c_3^2-h_1 s_3^2}{\sqrt{H_1(c_2^2-h_1 s_2^2)}}\left(dx+\frac{K_{\1}}{c_3^2-h_1 s_3^2}\right)^2,\cr
K_{\1}&=&s_3c_3(1-h_1)dt-\left(s_3c_2h_2-\frac{2\mu ac_3\sqrt{s_1c_1}s_2}{\rho^2}
\right)\frac{\sigma_3}{2},\cr
B_{\2}&=&-\frac{\mu s_1c_1}{2}(1+\cos\theta) d\varphi\wedge d\psi+\frac{s_2c_2(1-h_1)}{c_2^2-h_1s_2^2}dt\wedge dx\cr
&&\!\!\!\!\!+\frac{1}{2}\left(\frac{s_2c_3h_2}{c_2^2-h_1s_2^2}-\frac{2\mu a \sqrt{s_1c_1}c_2s_3}{\rho^2(c_2^2-h_1s_2^2)}\right)dx\wedge\sigma_3
+\frac{1}{2}\left(\frac{s_2s_3h_2}{c_2^2-h_1s_2^2}-\frac{2\mu a \sqrt{s_1c_1}c_2c_3}{\rho^2(c_2^2-h_1s_2^2)}\right)dt\wedge\sigma_3\,,\nn\\
A_{\1}&=&-k\cos\theta d\varphi\,,\qquad
\varphi=\log{\frac{c_2^2-h_1s_2^2}{H_1}}\,,\label{6D four charges 0}
\eea
where $(h_1,h_2, \Delta_\rho, H_1, \xi_1,\xi_2)$ are given by \eqref{xi1xi2}.

\subsection{Black Hole thermodynamics}

The boosted rotating dyonic string we constructed above is non-extremal, with a horizon located at the largest root $\rho_h$ of $\Delta_\rho=0$, where we can derive the temperature and entropy straightforwardly. The solution contains independent parameters $(\mu, a, \delta_1, \delta_2, \delta_3, k)$, giving rise to five independent conserved ``charge'' quantities: the mass $M$, the angular momentum $J_a=J_b$, electric $Q_e$ and magnetic $Q_m$ string charges associated with the 3-form field strength, the dipole charge $Q_D$ associated with the 2-form Maxwell field strength, and finally the boosted linear momentum $P_x$ along the string direction $x$. We find that the complete set thermodynamic variables are
\bea
M&=&\frac{\Omega _3}{8 \pi G_6}\frac{\mu  \xi _1}{\sqrt{\xi _2}}
(s_2^2+c_2^2+s_3^2+c_3^2)\,,\quad
 T=\frac{(\rho _h^2-2 a^2) \sqrt{\rho _h^2-a^2}}{2 \pi  \sqrt{\xi _2} \sqrt{s_1c_1}\rho _h^2 \big[(\rho _h^2-a^2)c_{23} +a^2 s_{23}\big]}\,,\cr
V_x&=&-\frac{(\rho _h^2-a^2) c_2 s_3 +a^2  s_2 c_3}{(\rho _h^2-a^2)c_{23} +a^2 s_{23}}\,,
\qquad \Omega_a=\Omega_b=\frac{2 a (\rho _h^2-a^2)}{\sqrt{s_1c_1}\rho _h^2 \big[(\rho _h^2-a^2)c_{23} +a^2 s_{23}\big]}\,,\cr
P_x&=&-\frac{\Omega _3}{4 \pi G_6 }\frac{\mu  \xi _1}{\sqrt{\xi _2}}s_3c_3\,,\qquad
J_a=J_b=\frac{\Omega _3}{8 \pi G_6}\frac{\mu a \xi _1}{\sqrt{\xi _2}}\sqrt{s_1c_1}(c_{23}-s_{23})\,,\cr
S&=&\frac{\Omega _3\xi _1  \rho _h^4}{4 (\rho _h^2-a^2)^{3/2}G_6}\sqrt{s_1c_1}\big[
 (\rho _h^2-a^2)c_{23}+a^2 s_{23}\big]\,,\cr
Q_e&=&\frac{\Omega _3}{4 \pi G_6}\frac{\mu  \xi _1}{\sqrt{\xi _2}}s_2c_2\,,\qquad
Q_m=\frac{\Omega _3}{4 \pi G_6}\mu s_1c_1\,,\qquad Q_D=\frac{\Omega_3}{2\pi G_6}k,\cr
\Phi_{e}&=&\frac{(\rho _h^2-a^2) s_2c_3  +a^2  c_2s_3 }{(\rho _h^2-a^2)c_{23} +a^2 s_{23}}\,,\qquad \Phi_D = -\fft{\sqrt{\xi_2} (\rho_h^2-2a^2)}{2 \xi_1 \rho_h^2 s_1 c_1}k\,,\qquad \Omega_3=2\pi^2\,.\label{thermo1}
\eea
All the thermodynamic quantities, except for $(M, \Phi_D)$ are obtained in the standard way, with no particular subtlety. (See e.g.~\cite{Townsend:1997ku} for a pedagogical review on the subject.) Since the metric is not asymptotic to flat spacetime, we do not have an independent way of computing the mass. We derive the formula of mass by requiring that the first law exists. Another subtlety is that, as explained earlier, the magnetic string charge $Q_m$ should be treated as a thermodynamic constant, {\it i.e.}, $\delta Q_m =0$.  We further require that $\delta Q_D=0$, influenced by the BPS condition studied later.  This allows to show, quite nontrivially, that the first law works with the mass derived. We then relax the condition $\delta Q_D=0$ and determine its potential $\Phi_D$. It is important to note that $\Phi_D$ does not depend on $(\delta_2,\delta_3)$. It is now straightforward to verify that the first law and the Smarr relation (e.g.~\cite{Townsend:1997ku}) are
\bea
&&\delta M=T\delta S+\Phi_e\delta Q_e + \Phi_D \delta Q_D+V_x\delta P_x+\Omega_a\delta J_a+\Omega_b\delta J_b\,,\cr
&&M-TS-\Phi_e Q_e-V_xP_x-\Omega_aJ_a-\Omega_bJ_b=0\,,
\eea
with the understanding that $\delta Q_m=0$. Note that the thermodynamic pair $(Q_D, \Phi_D)$ does not appear in the Smarr relation, consistent with the fact that $Q_D$ is dimensionless.

\subsection{A globally different solution}
\label{sec:sol2}

The equations of motion of Salam-Sezgin model is invariant under the trombone-like global symmetry
\be
g_{\mu\nu} \rightarrow \lambda g_{\mu\nu}\,,\qquad \varphi\rightarrow \varphi + 2 \log\lambda\,,\label{trombone}
\ee
since it has a consequence of uniformly scaling the whole Lagrangian. We can thus make the Lagrangian invariant by scaling the Newton's constant appropriately. Specifically, we consider
\bea
&&g_{\mu\nu}\rightarrow \sqrt{\frac{c_1}{s_1}}g_{\mu\nu}\,,
\qquad \varphi\rightarrow\varphi+\log\frac{c_1}{s_1}\,,\qquad \frac{1}{G_6}\rightarrow \frac{s_1}{c_1}\frac{1}{G_6}\,,\cr
&& t\rightarrow\sqrt{\frac{s_1}{c_1}}t\,,\qquad x\rightarrow\sqrt{\frac{s_1}{c_1}}x\,,
\eea
with the rest of fields and coordinates fixed. Use this symmetry, we can rewrite the seed solution \eqref{seed1} in a new way:
\bea
ds^2_{6,\mathrm{E}}&=&\frac{1}{\sqrt{H_1}}\left(-h_1dt^2+dx^2-h_2dt\sigma_3\right)
+\sqrt{H_1}\Big(\xi_2\frac{d\rho^2}{\Delta_\rho}+\frac{1}{4}\rho^2
\left(\sigma_3^2+\xi_1d\Omega_2^2\right)
\Big),\cr
A_{\1}&=&-k\cos\theta d\varphi,\quad B_{\2}=-\frac{\mu s_1c_1}{2}(1+\cos\theta) d\varphi\wedge d\psi-\frac{\mu a s_1}{\rho^2}dt\wedge\sigma_3,\cr
e^{-\varphi}&=&\fft{s_1^2}{c_1^2} H_1\,,\qquad H_1 =\fft{2\mu c_1^2}{\rho^2}\,,\qquad h_2=\frac{2\mu ac_1}{\rho^2}\,.
\label{seed2}
\eea
The remainder of the functions $(h_1, \Delta_\rho, H_1, \xi_1,\xi_2)$ are given by \eqref{xi1xi2}. After the same solution-generating process, we have a new boosted rotating dyonic string
\bea
ds_6^2&=&\frac{1}{\sqrt{H_1(c_2^2-h_1 s_2^2)c_1s_1^3}}\Bigg[
\frac{c_3^2 s_1^2 }{s_2^2}\left(c_2^2-h_1 s_2^2\right)dt^2-\frac{K_{\1}^2}{c_1^2 c_3^2-h_1 s_1^2 s_3^2}
-s_1^2 \left(\frac{c_2 c_3}{s_2}dt+h_2  s_2\frac{\sigma _3}{2} \right)^2\cr
&&
+c_1^2 \left(s_3dt+\frac{ \mu a s_1 s_2}{\rho^2} \sigma _3\right)^2
\Bigg]
+\sqrt{\frac{H_1(c_2^2-h_1 s_2^2) s_1}{c_1}}\left(
\xi_2\frac{d\rho^2}{\Delta_\rho}+\frac{1}{4}\rho^2\left(\sigma_3^2
+\xi_1d\Omega_2^2\right)\right)\cr
&&
+\frac{c_1^2 c_3^2-h_1 s_1^2 s_3^2}{\sqrt{ H_1 (c_2^2-h_1 s_2^2)c_1s_1^3 }}\left(dx+\frac{K_{\1}}{c_1^2c_3^2-h_1 s_1^2s_3^2}\right)^2,\cr
K_{\1}&=&s_3c_3(c_1^2-h_1s_1^2)dt-s_1\left(c_2 s_1 s_3h_2-\frac{2\mu ac_1^2 c_3 s_2}{\rho^2}
\right)\frac{\sigma_3}{2},\cr
B_{\2}&=&-\frac{\mu s_1c_1}{2}(1+\cos\theta) d\varphi\wedge d\psi+\frac{s_2c_2(1-h_1)}{c_2^2-h_1s_2^2}dt\wedge dx\cr
&&+\frac{1}{2}\left(\frac{s_2c_3h_2}{c_2^2-h_1s_2^2}-\frac{2\mu a s_1c_2s_3}{\rho^2(c_2^2-h_1s_2^2)}\right)dx\wedge\sigma_3
+\frac{1}{2}\left(\frac{s_2s_3h_2}{c_2^2-h_1s_2^2}-\frac{2\mu a s_1c_2c_3}{\rho^2(c_2^2-h_1s_2^2)}\right)dt\wedge\sigma_3,\cr
A_{\1}&=&-k\cos\theta d\varphi\,,\qquad
\varphi=\log{\left[\frac{c_2^2-h_1s_2^2}{H_1}\frac{c_1}{s_1}\right]}\,
,\qquad h_2=\frac{2\mu ac_1}{\rho^2}\,.\label{6D four charges}
\eea
Following the same strategy, we obtain the thermodynamical variables:
\bea
M&=&\frac{\Omega _3}{8 \pi G_6}\frac{\mu  \xi _1}{\sqrt{\xi _2}}
(s_1^2+c_1^2+s_2^2+c_2^2+s_3^2+c_3^2)\,,\quad
 T=\frac{(\rho _h^2-2 a^2) \sqrt{\rho _h^2-a^2}}{2 \pi  \sqrt{\xi _2} \rho _h^2 \big[(\rho _h^2-a^2)c_{123} +a^2 s_{123}\big]}\,,\cr
V_x&=&-\frac{(\rho _h^2-a^2)c_1 c_2 s_3 +a^2 s_1 s_2 c_3}{(\rho _h^2-a^2)c_{123} +a^2 s_{123}}\,,
\qquad \Omega_a=\Omega_b=\frac{2 a (\rho _h^2-a^2)}{\rho _h^2 \big[(\rho _h^2-a^2)c_{123} +a^2 s_{123}\big]},\cr
P_x&=&-\frac{\Omega _3}{4 \pi G_6 }\frac{\mu  \xi _1}{\sqrt{\xi _2}}s_3c_3\,,\qquad
J_a=J_b=\frac{\Omega _3}{8 \pi G_6}\frac{\mu a \xi _1}{\sqrt{\xi _2}}(c_{123}-s_{123}),\cr
S&=&\frac{\Omega _3\xi _1  \rho _h^4}{4 (\rho _h^2-a^2)^{3/2}G_6}\big[
 (\rho _h^2-a^2)c_{123}+a^2 s_{123}\big],\qquad Q_e=\frac{\Omega _3}{4 \pi G_6}\frac{\mu  \xi _1}{\sqrt{\xi _2}}s_2c_2,\cr
Q_m&=&\frac{\Omega _3}{4 \pi G_6}\mu s_1c_1,\qquad Q_D=\frac{\Omega_3}{2\pi G_6}k\,,\qquad \Phi_{e}=\frac{ (\rho _h^2-a^2)c_1 s_2 c_3+a^2  s_1c_2 s_3}{ (\rho _h^2-a^2)c_{123}+a^2 s_{123}}\,,\cr
\Phi_D &=& - \fft{g^2\sqrt{\xi_2}\,\rho_h^4}{\rho_h^2-a^2}(c_1^2+s_1^2) k +
\fft{\sqrt{\xi_2}\,k}{4\xi_1\,\rho_h^2 s_1 c_1}\Big(4a^2 + \rho_h^2(c_1^2 + s_1^2-2)\Big)\,.\label{thermo2}
\eea
The first law (with $\delta Q_m=0$) and the Smarr relation are
\bea
&&\delta M=T\delta S+\Phi_e\delta Q_e+ \Phi_D \delta Q_D+V_x\delta P_x+\Omega_a\delta J_a+\Omega_b\delta J_b,\cr
&&M=TS+\Phi_e Q_e +V_xP_x +\Omega_aJ_a +\Omega_bJ_b+\frac{(s_1^2+c_1^2)\xi_1}{2s_1c_1\sqrt{\xi_2}}Q_m.
\eea
We can see that in this new solution, the parameters $(\delta_1, \delta_2, \delta_3)$ enter the thermodynamic variables in a more symmetric manner, while in the old solution, the parameter $\delta_1$ stands out from $(\delta_2,\delta_3)$ parameters.

It should be pointed out that the symmetry \eqref{trombone} at the level equation of motion exists in ungauged supergravity; however, we do not apply this transformation on the asymptotically-flat string solutions since it can alter the asymptotic structure. In the gauged theory, the solutions are not asymptotically flat, and we do not have a fiducial spacetime. However, not all the scaling choices lead to a sensible description of black hole thermodynamics.  We only find one such alternate globally-different dyonic string that satisfies the first law of thermodynamics.

\section{Non-BPS and BPS extremal limits}

We have constructed two globally-different non-extremal boosted dyonic string solutions. We shall call \eqref{6D four charges 0} as solution A, and \eqref{6D four charges} as solution B. The horizon of both solution are determined as the largest root of the same function $\Delta_\rho$. The extremal limit corresponds to taking the temperature to zero. There are two ways of achieving this. One is simply set $\rho_h=\sqrt2 a$. In this case, $\mu=2a^2$ and $\Delta_\rho$ has double roots.  This is a rather standard and straightforward extremal limit, and we shall not discuss this further.

The other is the BPS limit where the parameters $\delta_i$ are sent to infinity while $(\mu,a)$ are sent to zeros such that charges and angular momentum remains finite and nonzero.  We shall discuss this BPS extremal limit in more detail. Specifically, the limit is achieved by taking
\be
\mu\sim\varepsilon,\quad e^{\delta_1}\sim\varepsilon^{-\frac{1}{2}},\quad e^{\delta_2}\sim\varepsilon^{\frac{1}{2}},\quad e^{\delta_3}\sim\varepsilon^{-\frac{1}{2}},\quad a\sim\varepsilon^{\frac{1}{2}}
\ee
and then sending $\varepsilon$ to zero while keeping the following parameters finite
\bea
\hbox{Solution A}:&&\qquad 2\mu s_i^2=q_i\,,\qquad \mu a\sqrt{s_1c_1}(c_{23}-s_{23})=-\frac{j}{2}\,,\cr
\hbox{Solution B}:&&\qquad 2\mu s_i^2=q_i\,,\qquad \mu a(c_{123}-s_{123})=-\frac{j}{2}\,.
\eea
In this limit, we find that both non-extremal dyonic solutions become the same and it is given by
\bea
ds_6^2&=&(H_PH_Q)^{-\frac{1}{2}}\Big[
-H_V^{-1}(dt+H_J\sigma_3)^2+H_V(dx-H_V^{-1}dt-H_V^{-1}H_J\sigma_3)^2
\Big]\cr
&&+(H_PH_Q)^{\frac{1}{2}}\Big[
\xi_2d\rho^2+\frac{\rho^2}{4}(\sigma_3+\xi_1d\Omega_2^2)
\Big],\cr
B_{\2}&=&(H_Q^{-1}-1)dt\wedge dx-\frac{H_J}{H_Q}dx\wedge \sigma_3-\frac{q_1}{4}(1+\cos\theta)d\varphi\wedge d\psi\,,\cr
A_{\1}&=&-k\cos\theta d\varphi\,,\qquad
e^\varphi=\frac{H_Q}{H_P}\,,\qquad
H_J=-\frac{j}{2\rho^2}\,,\quad H_Q=1+\frac{q_2}{\rho^2}\,,\qquad H_P=\frac{q_1}{\rho^2}\,,\cr
\xi_1&=&1+\frac{2k^2}{q_1}\,,\qquad \xi_2^{-1}=\xi_1^{-1}-2g^2q_1\,,
\eea
with
\bea
\hbox{Solution A}:\quad
H_V=1+\frac{q_3}{\rho^2}\,;\qquad
\hbox{Solution B}:\quad
H_V=1+\frac{q_3}{q_1}+\frac{q_3}{\rho^2}\,.
\eea
It is clear that $H_V$ is associated with the PP-wave component of the solution, arriving from the BPS limit of the Lorentz boost. Although the two $H_V$'s appear to be different, but the difference is trivial; a coordinate transformation $t\rightarrow t + q_3/(2q_1)$ renders them the same. In terms of the new variables, the charges become
\bea
&&Q_m=\frac{\Omega_3}{8\pi G_6}q_1\,,\qquad Q_e=-\frac{\Omega_3}{8\pi G_6}\frac{\xi_1}{\sqrt{\xi_2}}q_2\,,\qquad Q_D=\frac{\Omega_3}{2\pi G_6}k,\cr
&&P_x=-\frac{\Omega_3}{8\pi G_6}\frac{\xi_1}{\sqrt{\xi_2}}q_3\,,\qquad
J_a=J_b=-\frac{\Omega_3}{16\pi G_6}\frac{\xi_1}{\sqrt{\xi_2}}j\,.
\eea
Note that the horizon of the extremal solution is located at $\rho=0$, giving rise to the near-horizon geometry as a direct product of the boosted AdS$_3$ (or BTZ) metric and squashed $S^3$. From the volume of the squashed $S^3$, we obtain the entropy in the extremal limit:
\be
S=\frac{\Omega_3}{4 G_6}\xi_1\sqrt{q_1q_2q_3-j^2}\,.
\ee
This entropy formula is analogous to that of the extremal rotating black hole in five-dimensional ungauged STU model \cite{Cvetic:1996xz,Cvetic:1998xh}. The extra contribution from the magnetic dipole charge enters to the entropy formula as an overall factor via $\xi_1$.

The BPS limit on \eqref{thermo1} and \eqref{thermo2} leads to the following same thermodynamic potentials:
\be
\Phi_e=-1\,,\qquad V_x=-1\,,\qquad \Omega_a=\Omega_b=0\,.
\ee
The mass and $\Phi_D$ obtained from \eqref{thermo1} and \eqref{thermo2} are somewhat different. The BPS limit of \eqref{thermo1} gives
\be
M=-Q_e - P_x\,,\qquad \Phi_D=0\,.
\ee
The limit of \eqref{thermo2} gives
\be
M=\frac{\xi_1}{\sqrt{\xi_2}} Q_m-Q_e-P_x\,,\qquad
\Phi_D=\fft{k\sqrt{\xi_2}}{2\xi_1} (1-4g^2 q_1 \xi_1)\,.
\ee
However, the first law at this zero temperature, with $\delta Q_m=0$ is satisfied for both set of thermodynamic quantities, namely
\be
\delta M=\Phi_e\delta Q_e+V_x\delta P_x+\Omega_a\delta J_a+\Omega_b\delta J_b+  \Phi_D \delta Q_D\,.
\ee

The linear relation between the mass and charge suggests that the limits are BPS. In order to verify this, we need to construct Killing spinors, which satisfy three Killing spinor equations \cite{Salam:1984cj}
\bea
&&(\nabla_\mu - {\rm i} g A_\mu + \ft1{48} e^{\fft12\varphi} H^+_{\alpha\beta\gamma} \Gamma^{\alpha\beta\gamma} \Gamma_\mu)\epsilon=0\,,\cr
&&(\Gamma^\mu \partial_\mu \varphi - \ft16 e^{\fft12\varphi} H^-_{\mu\nu\rho} \Gamma^{\mu\nu\rho}) \epsilon=0\,,\qquad (e^{\fft14\varphi} F_{\mu\nu}\Gamma^{\mu\nu} - 8{\rm i} g e^{-\fft14 \varphi})\epsilon=0\,.
\eea
We write the dyonic string metric in the following vielbein basis
\be
ds^2=-a_1^2(dt+b_1\sigma_3)^2+a_2^2(dx+b_2dt+b_3\sigma_3)^2
+a_3^2(\sigma_1^2+\sigma_2^2)+a_4^2\sigma_3^2+a_5^2d\rho^2,
\ee
where $\sigma_3$ was given earlier and $\sigma_1$ and $\sigma_2$ are
\be
\sigma_1=\cos\psi d\theta+\sin\psi\sin\theta d\varphi\,,\qquad
\sigma_2=-\sin\psi d\theta-\cos\psi\sin\theta d\varphi\,.
\ee
Note that the round $S^2$ metric is $d\Omega_2^2=\sigma_1^2+ \sigma_2^2$. The vielbein are chosen to be
\bea
&&e^0=-a_1 (dt + b_1 \sigma_3)\,,\qquad e^1=a_2 (dx+b_2dt+b_3\sigma_3)\,,\qquad
e^2=a_1 \sigma_2\,,\cr
&& e^3=a_3 \sigma_2\,,\qquad e^4=-a_4 \sigma_3\,,\qquad e^5=a_5 d\rho\,.
\eea
We have chosen the chiral condition on the spinors to be
\be
(\Gamma^7+1)\epsilon=0\,.
\ee
We further find that the existence of Killing spinors requires the condition that relates the magnetic string and dipole charges, namely
\be
\xi_1^2=\xi_2,\qquad \Rightarrow\qquad g=\frac{k}{2k^2+q_1}\,.
\ee
In other words, the existence of Killing spinors require that the $Q_m$ and $Q_D$ charges are algebraically related, thereby reproducing the relation first given in \cite{Gueven:2003uw}. This particularly implies that $Q_D$ should not be treated as a thermodynamic variable in the BPS limit also, since we always imposed $\delta Q_m=0$.

The Killing spinor is then given by
\be
\epsilon=\frac{(q_3+\rho^2)^{\frac{1}{4}}}
{(q_2+\rho^2)^{\frac{1}{8}}}\epsilon_0\,,
\ee
where $\epsilon_0$ is a constant spinor, satifying the projection condition
\be
(\gamma^{23}-{\rm i})\epsilon=0\,,\qquad (\gamma^{2345}-1)\epsilon=0\,.
\ee
Note that the first projection arises from the magnetic dipole charge and the second projection arises from the magnetic string charge. Adding an electric string charge of appropriate sign does not affect the Killing spinor.  The PP-wave component would add a projection of $(\gamma^{01}+1)\epsilon=0$, which is automatically satisfied under the chiral projection. Thus the whole BPS solution preserves $1/4$ of supersymmetry and the rotation has no effect on the Killing spinors.

\section{Conclusions}

In this paper, we showed that despite the $U(1)_R$ symmetry gauging, the Salam-Sezgin model still has a global $T$-duality symmetry upon a circle reduction. The symmetry acts nonlinearly on the $SO(2,1)/SO(2)$ scalar coset, but linearly on the three abelian vectors which form an $SO(2,1)$ triplet. The antisymmetric tensor is the singlet of $SO(2,1)$.

In our construction of string solutions, we first constructed a simpler rotating magnetic string as a seed solution, and derived the general boosted rotating dyonic strings by applying the $T$-duality symmetry. The general solution is characterized by the mass $M$, electric and magnetic string charges $(Q_e,Q_m)$, a magnetic dipole charge $Q_D$ and two equal angular momenta $J_a=J_b$, together with a linear momentum $P_x$ along the string direction. We analyzed the global structure and established the first law of thermodynamics. Owing to the fact that the solution are not asymptotically flat, we do not have a fiducial Minkowski spacetime. This allows us to use trombone symmetry to obtain a globally different dyonic string solution that satisfy a different version of the first law. Under the BPS limit we found that both non-extremal solutions reduces to the same solution. The existence of Killing spinor enforces an algebraic constraint on the magnetic string and the dipole charges, obtained in \cite{Gueven:2003uw}, but for the general non-BPS solutions, all charges are completely independent. It is worth pointing out that the seed solution carrying magnetic string and dipole charges preserves only $1/4$ in the BPS limit, i.e.~the minimum amount of supersymmetry. The proper inclusion of the electric string charge, the PP-wave momentum, or the angular momentum does not break the supersymmetry any further.

The reason we consider the two equal angular momentum case, i.e.~$J_a=J_b$ is that in the static case, the 3-sphere in the transverse space is already squashed to be a $U(1)$ bundle over $S^2$. It is not clear whether squashed 3-sphere allows to have a more general  $J_a\ne J_b$ rotations, but it certainly deserves further investigation. The Salam-Sezgin model by itself suffers from local anomalies. Its anomaly-free extensions \cite{Randjbar-Daemi:1985tdc,Avramis:2005hc,Avramis:2005qt} based on the Green-Schwarz mechanism necessarily introduce higher derivative terms in the effective action \cite{Pang:2020rir}, analogous to the well-studied heterotic supergravity in ten dimensions \cite{Bonora:1986ix, Sakai:1986hh,DAuria:1988vnx, Bellucci:1988ff,Bergshoeff:1988nn, Bellucci:1990fa}. It is thus interesting to see how these higher order interactions will modify the thermodynamic quantities of the solutions. This is more challenging than the case of heterotic supergravity compactified on K3, where the higher derivative corrections to the onshell action of black string \cite{Ma:2022gtm} can be obtained using the trick of \cite{Reall:2019sah} without actually solving the field equations. With the $U(1)_R$ symmetry gauging, the solution is no longer asymptotic to Minkowski space nor to AdS, and therefore it is unclear whether the trick, or its AdS improved versions \cite{Xiao:2023two,Hu:2023gru}, still applies.

\section*{Acknowledgement}

We are grateful to Shuang-Qing Wu for pointing out the typo in \eqref{HE} in the original version. This work is supported in part by the National Natural Science Foundation of China (NSFC) grants No.~11935009 and No.~12375052 and No.~12175164.  Y.P.~is also supported by the National Key Research and Development Program under grant No. 2022YFE0134300.

\end{document}